\documentclass[aps, twocolumn, superscriptaddress, showpacs, nofootinbib, longbibliography]{revtex4-1}

\usepackage[utf8]{inputenc}
\usepackage[T1]{fontenc}
\usepackage{ae,aecompl} 
\usepackage{graphicx}
\usepackage{amsmath}
\usepackage{xcolor}
\usepackage{amssymb}
\usepackage{latexsym}
\usepackage{wasysym}
\usepackage{psfrag}
\usepackage{ifthen}
\usepackage[citecolor=blue,colorlinks=true]{hyperref}
\usepackage{float}
\usepackage[utf8]{inputenc}
\usepackage{lineno}
\usepackage{multirow}
\usepackage{subcaption}
\usepackage{orcidlink}

\begin{document}

\title{Search for dynamical black hole captures enhanced with Gaussian mixture modeling}

\author{Leigh Smith\orcidlink{0000-0002-3035-0947}}
\affiliation{Dipartimento di Fisica, Università di Trieste, I-34127 Trieste, Italy}
\affiliation{INFN, Sezione di Trieste, I-34127 Trieste, Italy}
\affiliation{SUPA, School of Physics and Astronomy, University of Glasgow, Glasgow G12 8QQ, United Kingdom}
\author{Shubhanshu Tiwari\orcidlink{0000-0003-1611-6625}}
\affiliation{Physik-Institut, Universität Zürich, Winterthurerstrasse 190, 8057 Zürich}
\author{Michael Ebersold\orcidlink{0000-0003-4631-1771}}
\affiliation{Physik-Institut, Universität Zürich, Winterthurerstrasse 190, 8057 Zürich}
\author{Yeong-Bok Bae\orcidlink{0000-0003-3093-9206}}
\affiliation{Department of Physics, Chung-Ang University, Seoul 06974, Republic of Korea}
\author{Gungwon Kang}
\affiliation{Department of Physics, Chung-Ang University, Seoul 06974, Republic of Korea}
\author{Ik Siong Heng\orcidlink{0000-0002-1977-0019} }
\affiliation{SUPA, School of Physics and Astronomy, University of Glasgow, Glasgow G12 8QQ, United Kingdom}

\begin{abstract}

Gravitational waves (GWs) are expected to originate from black holes interacting dynamically in dense astrophysical environments. In such environments, given that the velocity and cross section between the interacting black holes is low, dynamical capture may occur. Such events merge on very short timescales with high eccentricities and are expected to be detectable in the LIGO-Virgo-KAGRA (LVK) sensitivity band. 
In this work, we present a dedicated search for dynamical black hole capture events in the third LVK observing run with the coherent WaveBurst (cWB) algorithm enhanced with Gaussian mixture modeling (GMM) post-production. With this we consider two applications of GMM: a weakly-modeled approach searching for generic short transients under minimal assumptions, and a population informed approach, in which the GMM model is provided information on the parameter space occupied by the capture population. Although our search does not find any new significant GW events, we find that an informed GMM approach brings significant sensitivity improvements, enabling the detection of dynamical capture events up to a distance of 1.9 Gpc for a 200 $M_{\odot}$ equal mass binary. We present updated upper limit estimates of the rate at 90\% confidence, the most stringent of which is 0.15 Gpc$^{-3}$yr$^{-1}$, a 34\% improvement with respect to previous observational estimates. Furthermore, while the weakly-modeled GMM approach is less sensitive to dynamical capture systems, we find that it is possible for these events to be detected up to a distance of 1 Gpc in the cWB-GMM all-sky short search under minimal assumptions. Finally, with the confident detection of GW190521, we estimate the rate of similar events to be 0.94 Gpc$^{-3}$yr$^{-1}$, assuming the event originated from a dynamical capture.

\end{abstract}

\maketitle

\section{Introduction}\label{Sec:intro}

A fraction of binary black holes are expected to be formed dynamically in dense astrophysical environments, like the galactic center or globular clusters \cite{Mapelli:2020BBH_form}. In these environments, there can be gravitational interactions of black holes leading to either hyperbolic encounters or dynamical capture, occasionally referred to as radiation-driven capture. In the interactions leading to capture two scenarios can occur: either the captured black holes are still far apart and will evolve for a long time before merging, in which the circularization of their orbits will occur, or the black holes are close by and merge quickly after the capture, retaining an eccentric orbit \cite{Bae:2017capture}. 
In such scenarios, gravitational waves (GWs) are emitted in a form where the inspiral phase is absent, and the waveform transitions directly to the merger and ringdown phases. It is expected for a fraction of such rapidly-merging captures to lie within the sensitivity band of the LIGO-Virgo-KAGRA (LVK) detectors. In particular, a recent study has suggested that GW190521 may have originated from a dynamical capture event, highlighting the growing interest in the detection of GW signals of this nature \cite{Gamba:2023capture}. 

It is possible for dedicated searches for compact binary coalescences (CBCs) utilising matched filtering \cite{LIGOScientific:matchedfilter} to detect a fraction of such dynamical capture events, however it is not guaranteed that their short duration, highly eccentric waveforms are fully covered by template banks. Instead, one can consider weakly-modelled search methods that rely on the identification of excess coherent energy across a network of detectors.
Weakly-modelled search algorithms are sensitive to GW transients from a plethora of astrophysical sources, such as core-collapse supernovae (CCSN) \cite{CCSN-o3-Szczepanczyk,Szczepanczyk:2021adv-ccsn}, non-linear memory effects \cite{Ebersold:2020nl-mem}, cosmic strings \cite{LIGOScientific:2021cso3}, neutron star glitches \cite{Lopez:2022nsglitch,Yim:2020nsglitch} and a range of CBC systems, in particular intermediate-mass black hole (IMBH) binaries \cite{o3-imbh} and eccentric black hole binaries (eBBH) \cite{o3-ebbh}. 
Dynamical captures of black holes are also detectable with such weakly-modeled searches, with the expected signal morphology lying directly within the parameters of the all-sky short low-frequency search, which searches for GW transients with duration 0.1 ms to 10 s with frequency range 32 Hz to 1024 Hz \cite{allsky-o1,allsky-o2,allskyo3}.

Our previous work \cite{Ebersold:2022capture} investigated the sensitivity to dynamical black hole capture events during the third LVK observing run (O3) with the coherent WaveBurst (cWB) algorithm \cite{Klimenko_2008cwb,Klimenko:2015cwb,Drago:2020cwb}, finding that such events may be detected up to 1.64 Gpc, and placing the most stringent rate upper limit at 90\% confidence as 0.23 Gpc$^{-3}$yr$^{-1}$. 
Although good sensitivity was found with this search, a fraction of the waveforms were affected by a form of transient detector noise, called blip glitches \cite{Cabero:2019blip}. 
The post-processing in this algorithm utilized selection cuts and binning on signal attributes to distinguish GW signals from noise artifacts, however it has recently been shown that the implementation of machine learning techniques as post-processing to the cWB search can significantly improve sensitivities, particularly in mitigating the affect of blip glitches \cite{Smith:2024gmm,xgboost_2023,Ghosh:2025gmm-jsd}. 

In this work, we explore the sensitivity to dynamical capture events through the Gaussian mixture modeling (GMM) post-production approach to the cWB pipeline \cite{Gayathri2020gmm,Lopez2022gmm,Smith:2024gmm} in an attempt to detect GWs from dynamical black hole capture. GMM allows for multidimensional non-Gaussian data to be represented as a superposition of multivariate Gaussians. It can be applied as a detection tool by representing the background and signal populations of GW searches as separate multiple-Gaussian models and constructing a statistic which ranks events on the likelihood of belonging to each population. In recent work \cite{Smith:2024gmm}, it was demonstrated that the application of the cWB-GMM algorithm to the search for all-sky, short duration GW transients is capable of bringing significant sensitivity improvements. For this work we consider two GMM approaches: the generic weakly-modeled approach assuming no signal morphology as applied in \cite{Smith:2024gmm}, and the first time application of an `informed-GMM', in which the GMM signal model is informed of the parameter space of the dynamical capture population in an attempt to further enhance sensitivities. 
With both GMM approaches, we calculate the sensitive distance, comparing the results with the previous sensitivity study \cite{Ebersold:2022capture}. With significantly enhanced sensitivity being observed with the informed-GMM post-production, we present search results for dynamical capture during O3, utilizing the results to calculate updated observational rate upper limits.

The paper is structured as follows: Section \ref{Sec:Capture} provides details of the dynamical black hole capture waveforms utilized within this study. Section \ref{Sec:Methodology} details the cWB+GMM algorithm, providing details on the two approaches of GMM and the O3 data utilized in the sensitivity study. In Section \ref{Sec:Results} we present the results of the search and sensitivity study, including sensitive distance and rate upper limit calculations. Finally, we discuss our concluding remarks and scope for future improvements in Section \ref{Sec:conclusions}.

\section{Dynamical Capture waveforms}\label{Sec:Capture}

Dynamical capture refers to the process by which two initially unbound black holes become bound through close interactions, forming an eccentric orbit that eventually leads to a merger. The nature of the orbit depends on the initial angular momentum $L$ of the system. For large $L$, the orbit follows a fly-by trajectory and the black holes remain unbound. However, for small $L$, the system becomes bound and transitions to a dynamical capture. For capture events, as $L$ increases towards the boundary between capture and fly-by, the system evolves from direct mergers to those involving an increasing number of cycles before merger. Close to this boundary, black holes may undergo multiple close encounters before eventually merging.
In this study, we focus on GW detection only from dynamical capture orbits, focusing on a subset of systems which either directly merge or have at most one encounter before merger, referred to as having `double blip' structure. 
In such systems, the merger occurs on short timescales, reserving the high eccentricity of the orbit. 

We employ waveforms from non-spinning black holes on parabolic orbits, obtained from numerical relativity simulations across a range of mass ratios and angular momentum configurations, as described in \cite{Bae:2017capture}. The assumption of a parabolic orbit implies that the initial orbital energy is zero, thus simplifying the simulations. 
In most astrophysical scenarios, dynamical capture occurs with relatively low orbital energy, close to zero, thus nearly at the parabolic limit \cite{East:2012}. Furthermore, the orbital configuration near the pericenter - where most of the gravitational wave emission occurs - or just before merger can often be well approximated by a parabolic orbit. These considerations justify the use of waveforms from parabolic orbits \cite{Quinlan:1989parabola, Bae:2017capture}.

\begin{figure*}
    \centering
    \includegraphics[width=\textwidth]{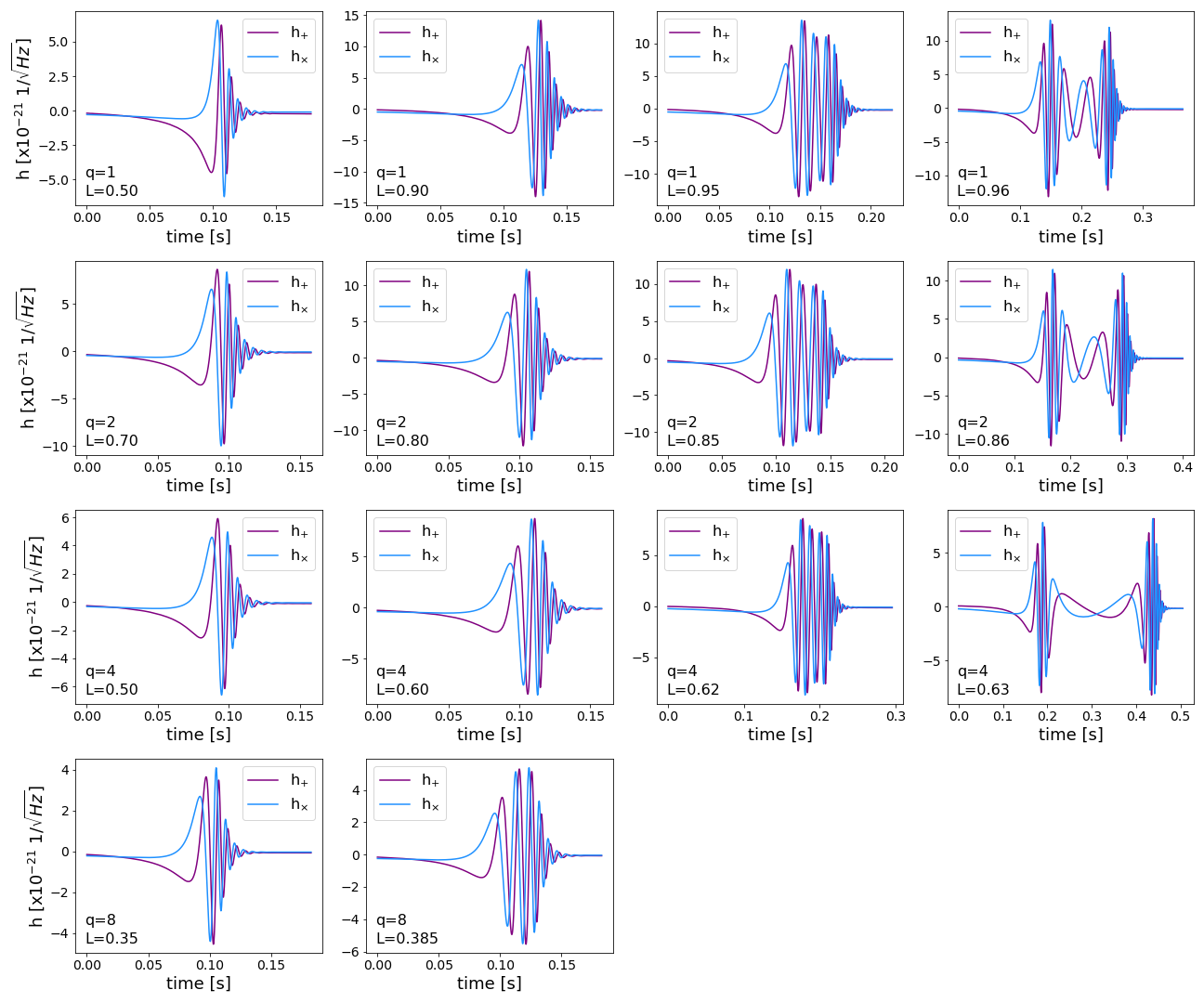}
    \caption{ Example of the considered set of dynamical capture waveforms from Bae et al. \cite{Bae:2017capture} with total mass 100 $M_{\odot}$ at a distance of 100 Mpc. Waveforms vary over mass ratio $q=[1,2,4,8]$ (increasing downwards), and initial angular momentum (increasing left to right). Note that only two initial angular momentum configurations are considered for $q=8$ due to a lower initial angular momentum threshold between dynamical capture and fly-by. }
    \label{fig:capture_waveforms}
\end{figure*}

The set of capture waveforms considered consists of 14 simulations distributed over mass ratio $q=[1,2,4,8]$ and initial angular momentum $L$. The values of $L$ are dependent on the threshold between the capture and fly-by orbits, which varies with system parameters such as the mass ratio.
An example of these simulations scaled to have a total mass $M_{tot}$ = 100 $M_{\odot}$ can be seen in Figure \ref{fig:capture_waveforms}.

\section{Methodology}\label{Sec:Methodology}

\subsection{Coherent WaveBurst}

The coherent WaveBurst (cWB) algorithm \cite{Klimenko_2008cwb,Klimenko:2015cwb,Drago:2020cwb,Klimenko:2022wavescan} searches for GW signals while making minimal assumptions on a source's morphology, sky direction and polarization, instead searching for excess coherent energy across a network of detectors. cWB is used for a wide range of GW transient searches, including both model-agnostic all-sky searches \cite{allskyo3,o3-allskylong} and model-informed binary black hole searches \cite{o3-imbh,o3-ebbh,Mishra:2024XPO3}.

cWB converts strain data from the detectors into the time-frequency domain via the Wilson-Daucechiers-Meyer (WDM) wavelet transformation \cite{Necula:2012wdm}, and clusters coherent time-frequency pixels across the network. A range of attributes are calculated for the clusters, quantifying features such as coherence across the network and energy content of the event. The clusters will be reconstructed via maximized likelihood and labeled as possible GW signals, or {\it triggers}, if they pass thresholds on coherent energy. The remaining range of trigger attributes can be utilized to distinguish real signals from occurrences of transient noise, or {\it glitches}.  
Background noise is estimated by time-shifting data between detectors by an amount that is not physical for a GW to travel between them. Such a data set only consists of noise and detector glitches, and is utilized for the calculation of false alarm rate (FAR), which is used as a significance measure. 

In the all-sky search for short transients in O3 data \cite{allskyo3}, cWB triggers were post-processed to distinguish signals from glitches through the manipulation of statistical attributes and application of selection cuts. Subsequently, triggers were split into 3 post-production bins based on their {\it Qveto} estimates, which quantifies the similarity of trigger morphology to the known population of blip glitches \cite{Cabero:2019blip}. 
While such a method aids the detection of signals with high Qveto (lying in the `clean' bins), the low Qveto bin is often plagued with blip glitches, which makes the detection of signals within this bin difficult. This was the post-production methodology considered in our previous sensitivity study for dynamical captures \cite{Ebersold:2022capture}, causing a lowered sensitivity to short duration, low $L$ waveforms. Below we present an alternative post-processing method which enhances the classification of GWs against glitches for cWB.

\subsection{Gaussian Mixture Modeling as post-production}
\label{Sec:GMM}

In recent works \cite{Gayathri2020gmm,Lopez2022gmm,Smith:2024gmm} it has been demonstrated that the supervised Machine Learning based method of Gaussian mixture modeling (GMM) can be employed as an effective technique for post-production to the cWB algorithm, improving the sensitivity to short duration GW transients and mitigating the effect of blip glitches. GMM allows for multi-modal distributions within a multidimensional space to be represented as a superposition of Gaussians. In the context of cWB, GMM is applied to triggers using a selection of their statistical attributes to model both the noise and signal populations as superpositions of Gaussians over the multi-dimensional attribute space. The choice of statistical attributes should represent the noise and signal populations well, and are detailed in Appendix \ref{app:attributes}. The number of Gaussians per model is a hyper-parameter of the method, selected through the optimization of likelihood on some validation data \cite{Smith:2024gmm}. Once the number of Gaussians are selected, models are trained by optimizing the individual Gaussian weight, covariance and mean parameters with the Expectation Maximization algorithm \cite{EM_1977}. 

From the two distinct models representing the signal and noise populations, we can construct a likelihood ratio statistic, defined as the GMM detection statistic, $T$:
\begin{equation}
    T = W_s - W_n
\end{equation}
for $W_s, W_n$ the likelihood of a given event belonging to the signal model or noise model, respectively. Such a statistic distinguishes GW signals from glitches by promoting probable signals to regions of high $T$, and noise to regions of low $T$. Estimations of $T$ on the accumulated background allow for the FAR estimates to be calculated on possible GW signals.

In the cWB+GMM application to the O3 all-sky short search \cite{Smith:2024gmm}, the noise model is trained on a portion of time-shifted background triggers, while the signal model is trained on some generic distribution of white noise burst (WNB) simulations in order to ensure minimal assumptions on source morphology during training. While this approach is favored for model-agnostic searches, it is possible to construct a `population informed' GMM analysis simply through the selection of training data for the signal model. 
Here, we search for dynamical capture events through the application of cWB+GMM, building upon sensitivity estimates reported with cWB alone (referred to as {\it standard cWB}) \cite{Ebersold:2022capture}. We follow the most recent GMM methodology outlined in \cite{Smith:2024gmm}, considering the all-sky short generic approach for a weakly-modeled analysis, in addition to a population informed GMM approach explored for the first time to target the parameter distribution of dynamical capture events. Details of both GMM approaches are provided in Section \ref{Sec:search}.

\subsection{Sensitivity study}\label{Sec:search}

For the sensitivity study we consider data taken from the LVK third observing run (O3). The run consists of two epochs separated by a one-month commissioning period, with O3a running from 1st April 2019 to 1st October 2019 and O3b running from 1st November 2019 to 27th March 2020. We consider only the 2-detector LIGO-Hanford, LIGO-Livingston (HL) network, in which 104.9 days of coincident data is collected in O3a, and 93.4 days of coincident data in O3b. Time-shifted background data accumulates to a total of 980.7 years for O3a and 1096.0 years for O3b. 

Our work builds upon the previous dynamical black hole capture sensitivity study with standard cWB detailed in \cite{Ebersold:2022capture}, thus we consider the same set of capture simulations with the same injection parameters. Each of the 14 numerical relativity waveforms, which vary over mass ratio $q$ and initial angular momentum $L$ as detailed in Section \ref{Sec:Capture}, is scaled to four total mass values $M_{tot}=[20,50,100,200] M_{\odot}$, totaling to 56 waveforms in the injection set. These are injected uniformly in sky location and inclination angle, and uniformly distributed in co-moving volume up to a maximum redshift $z^i_{\text{max}}$. Maximum redshift is chosen individually for each waveform in the injection set so that no signals are recovered at maximum distance. 
Each signal is individually redshifted according to the cosmological parameters provided in \cite{Planck:2018}. For each waveform in the injection set over 200,000 injections are made into the data over the entirety of O3. 

We consider the use of cWB under `all-sky short, low frequency' search settings, meaning we search for transients in the [32,1024] Hz range, with duration up to $\sim10$ s. 
cWB is applied to the injection data, outputting recovered capture injections as triggers, which can be post-processed by GMM to optimize sensitivity. GMM models are trained separately for O3a and O3b, with results combined in the final stages. 
As stated above, we consider a generic weakly-modeled approach and an informed approach by varying the data utilized in training the signal model. 
For both the generic and informed approach, the same background data is considered for training the noise model, consisting of 70\% of the time-shifted background. 10\% is used for validation data, with the number of Gaussian components in the models being optimized separately for each approach. Finally, 20\% of the background data is reserved for testing, resulting in 196.19 years of background for FAR calculation in O3a and 219.2 years in O3b.

\subsubsection{Generic all-sky short model}

The weakly-modeled GMM approach utilizes signal training data detailed in \cite{Smith:2024gmm}. Since the objective here was to hold minimal assumptions on source morphology, the training data consists of generic WNB simulations with duration, bandwidth and frequency distributed over the parameter space of the search. 80\% of the simulated WNB data is utilized for model training, while the remaining 20\% is used for the optimization of Gaussian components in the models. 

Since this GMM training set-up is the typical approach utilized in the weakly-modeled search for short duration transients, we can investigate whether it is feasible to detect dynamical black hole capture events in the all-sky short transient search with cWB+GMM. 

\begin{figure*}
    \centering
    \includegraphics[width=\textwidth]{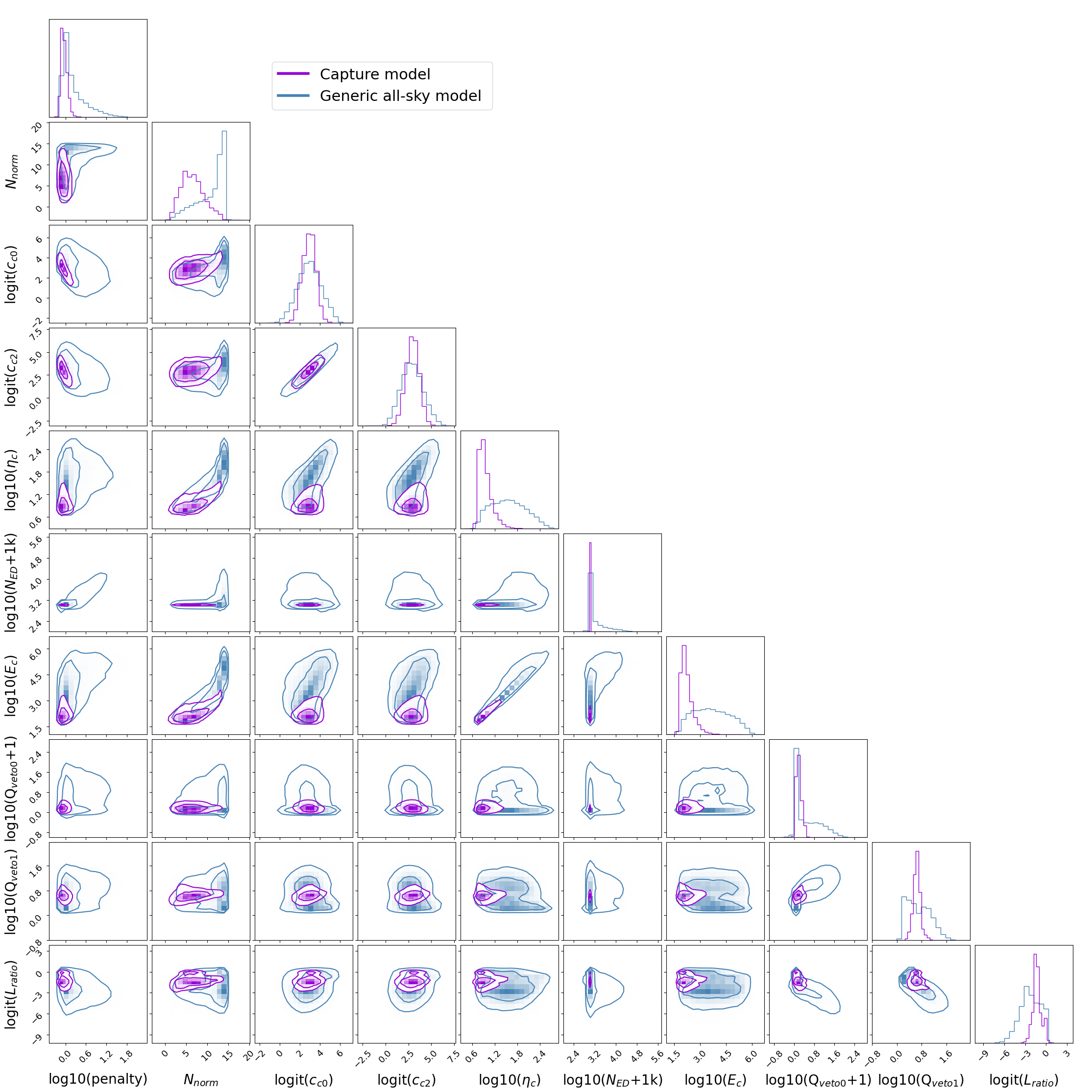}
    \caption{Corner plot of the generic all-sky short GMM signal model (blue) and the capture-informed GMM signal model (purple) across the selected cWB attribute space. It can be seen that the informed model is concentrated on a smaller region of the parameter space, where signals from capture events are expected to lie. 
    Further details of the selected cWB attributes can be seen in Appendix \ref{app:attributes}. }
    \label{fig:corner_allsky_cap}
\end{figure*}

\subsubsection{Capture-informed model}

For the capture-informed GMM approach, we train the signal model on a portion of the dynamical capture injections processed by cWB. 
For each injected total mass, we consider 10\% of triggers as validation data to optimize the number of Gaussians in the model and 20\% for testing sensitivities. The training data is formed from the remaining data, however due to cWB not recovering an equal number of triggers from each total mass, we must select the same number of triggers from each mass to avoid bias in the models. cWB recovers the least of 20 $M_{\odot}$ injections, since these have the lowest energy associated with them and hence the lowest strain amplitude. The remaining 70\% of 20 $M_{\odot}$ data contains $\sim$8,000 triggers, hence we take 8,000 triggers to be the size of training for all total masses. 
Once the data sets have been split for each total mass, they are combined so that a single model is constructed over all total masses. 

A comparison of the generic all-sky short signal model vs. the capture-informed signal model over the selected cWB attributes is shown in Figure \ref{fig:corner_allsky_cap}. While the generic all-sky model covers the region of the parameter space where dynamical capture signals lie, the informed model is more densely concentrated over the targeted region, and thus can be expected to boost the likelihood of capture events against the background population, further increasing sensitivity.

\section{Results}\label{Sec:Results}

%%%%%%%%%%%%%%%%%%% Figure for section below %%%%%%%%%%%%%%%%%%%%%%%%%%
\begin{figure*}
    \centering
     \begin{subfigure}{0.49\textwidth}
         \centering
         \includegraphics[width=\textwidth]{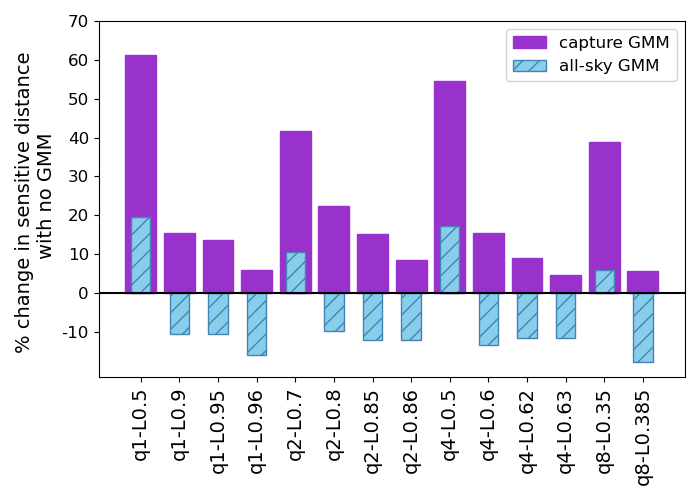}
         \caption{20$M_{\odot}$ total mass}
     \end{subfigure}
     \begin{subfigure}{0.49\textwidth}
         \centering
         \includegraphics[width=\textwidth]{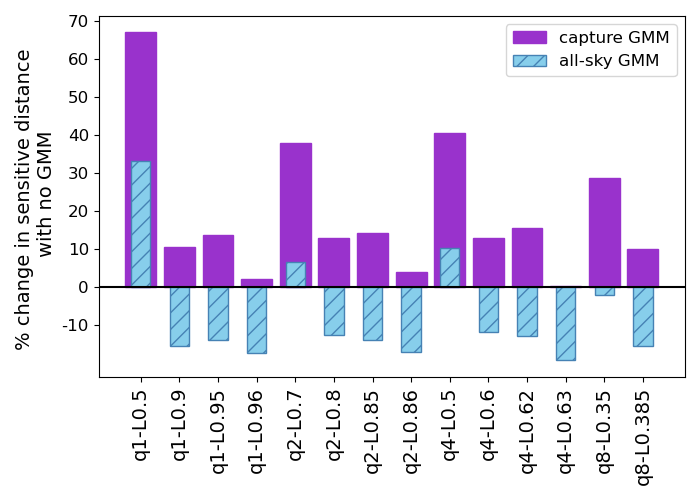}
         \caption{50$M_{\odot}$ total mass}
     \end{subfigure}
     \hfill
     \begin{subfigure}[b]{0.49\textwidth}
         \centering
         \includegraphics[width=\textwidth]{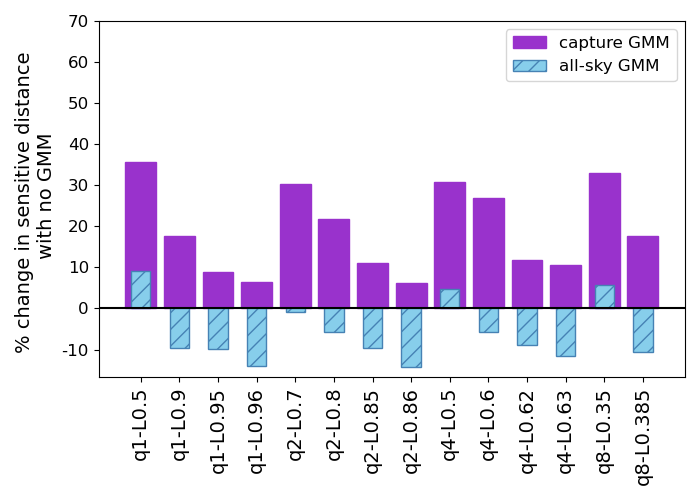}
         \caption{100$M_{\odot}$ total mass}
     \end{subfigure}
     \begin{subfigure}{0.49\textwidth}
         \centering
         \includegraphics[width=\textwidth]{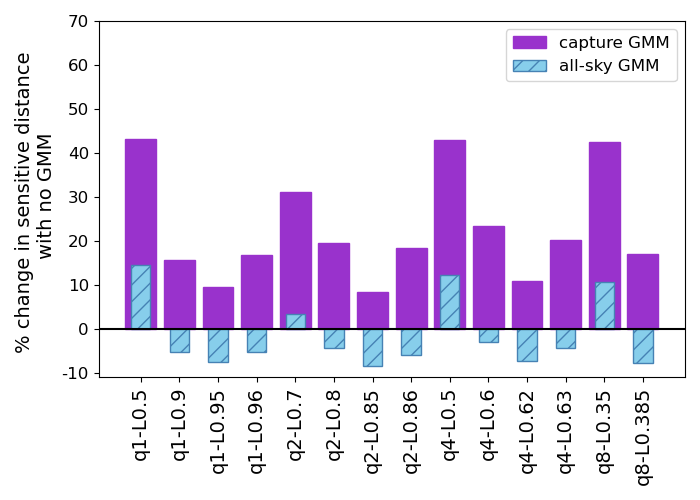}
         \caption{200$M_{\odot}$ total mass}
     \end{subfigure}
        \caption{Percentage change in sensitive distance of the two GMM approaches with respect to the previous study in \cite{Ebersold:2022capture} (no GMM post-production) at iFAR $\geq$ 100 years. Results for the all-sky short GMM are shown in blue, while results for the capture-informed GMM are in purple. Results are split by total mass value, with total mass 20$M_{\odot}$ shown in (a), 50$M_{\odot}$ in (b), 100$M_{\odot}$ in (c) and 200$M_{\odot}$ in (d). }
        \label{fig:percenatge_sens_dist}
\end{figure*}
%%%%%%%%%%%%%%%%%%%%%%%%%%%%%%%%%%%%%%%%%%%%%%%%%%%%

\subsection{Sensitive distance from simulations}

\begin{figure*}
    \centering
    \includegraphics[width=\textwidth]{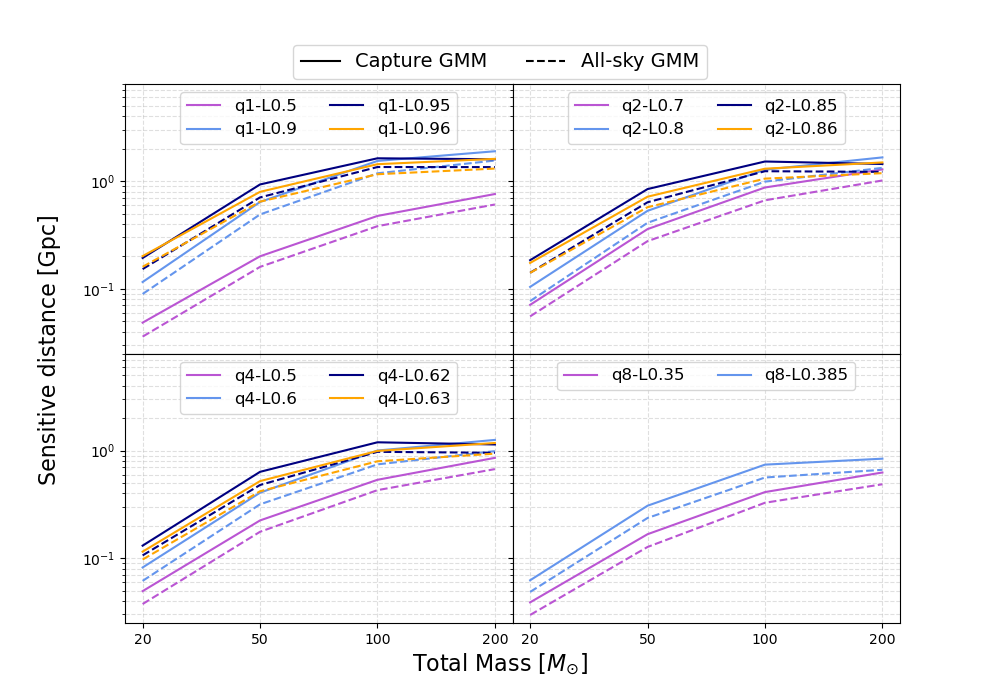}
    \caption{ Distribution of sensitive distance for each considered $q,L$ waveform over total mass values for iFAR $\geq$ 100 years. The generic all-sky short GMM analysis is represented by dashed lines, while the capture-informed GMM analysis is represented by sold lines. Results are split by mass ratio, increasing from top-left to bottom-right. }
    \label{fig:sens_dist}
\end{figure*}

We estimate the sensitivity of our analysis to dynamical capture events by introducing the sensitive distance measure. This is based upon the sensitive averaged spacetime-volume per waveform \cite{Tiwari:2017sens-dist,LIGOScientific:2016gw150914rate,LIGOScientific:2016gw150914rate-supp}:
\begin{equation}
    \langle VT \rangle =  \int  dz d\theta \frac{dV_c}{dz} \frac{1}{1+z} p_{pop}(\theta)f(z,\theta)T
\end{equation}
for differential co-moving volume $dV_c/dz$, distribution of binary parameters $p_{pop}$ and length of observation in the detector frame $T$. For our study, we consider the distribution of sky location to be uniform and inclination angle, $\iota$, to be distributed uniformly in $\cos(\iota)$. $f(z,\theta)$ is the probability of recovering an injected signal with parameters $\theta=[q,L,M_{tot}$] at redshift $z$.

From the averaged spacetime-volume the corresponding sensitive distance is computed via:
\begin{equation}
    D_{\langle VT \rangle} = \left( \frac{3 \langle VT \rangle}{4 \pi T_s} \right)^{1/3}
\end{equation}
for length of analysed data $T_s$. All results are presented above a given threshold of inverse false alarm (iFAR), a significance estimate defined as the reciprocal of false alarms with respect to the background distribution. 

First, we present the percentage change in sensitive distance at iFAR $\geq$ 100 years for each GMM approach with respect to the sensitive distance found with the standard cWB analysis \cite{Ebersold:2022capture}, to directly highlight the impact brought by each GMM method. These results are seen in Figure \ref{fig:percenatge_sens_dist}, separated by total mass. 
For the generic all-sky analysis, it is seen that an improvement in sensitive distance is only observed for low initial angular momentum waveforms. This improvement is due to the ability of GMM to mitigate the effect of blip glitches, which previously limited the sensitivity of low $L$ waveforms. For all other waveforms, a sensitivity loss of approximately 10\% is observed. Despite this, it is still possible to detect such waveforms in the all-sky short search, as discussed below. 

In the case of the informed GMM approach, significant sensitivity improvement is seen across all considered waveforms, with particular increase for low $L$ waveforms up to 65\%. In general, the sensitivity gains across all waveforms are expected, as our model targets the sub-set of the multidimensional attribute space that corresponds to capture waveforms, thereby allowing better differentiation between signal and noise. The particularly high sensitivity increase for low $L$ waveforms again arises due to the ability of GMM to mitigate the affect of blip glitches.

We also present the direct estimates of sensitive distance at iFAR $\geq$ 100 years for the generic all-sky and capture-informed GMM over total mass for each waveform in Figure \ref{fig:sens_dist}. The sensitive distance estimates are directly stated for both GMM approaches and the standard cWB analysis in Table \ref{tab:comparison}.

Although the largest sensitivity improvement is observed for low $L$ waveforms relative to the previous analysis, both GMM approaches remain the least sensitive to such waveforms across all total masses and mass ratios. 
The maximum sensitive distance is achieved for the $q=1, L=0.9 $, M$_{tot}=200M_{\odot}$ waveform with the informed GMM approach, reaching 1.9 Gpc. 
As expected, sensitive distance generally increases with $L$, since the most gravitational energy is radiated near the boundary between capture and fly-by systems. While this trend holds in most cases, a decrease in sensitivity is observed for systems with the highest $L$ values, likely due to their `double blip' structure, which may not always be well reconstructed by cWB. 
Sensitive distance is expected to increase with total mass, since the radiated gravitational energy increases with total mass. 

Finally, despite the 10\% reduction in sensitivity observed with the generic all-sky GMM approach, it can be seen that dynamical capture events are still detectable out to $\sim 1$ Gpc, suggesting it is feasible to detect such events in a typical weakly-modeled, low frequency search for short GW transients.

\subsection{Search results}

\begin{figure}
    \centering
    \includegraphics[width=\linewidth]{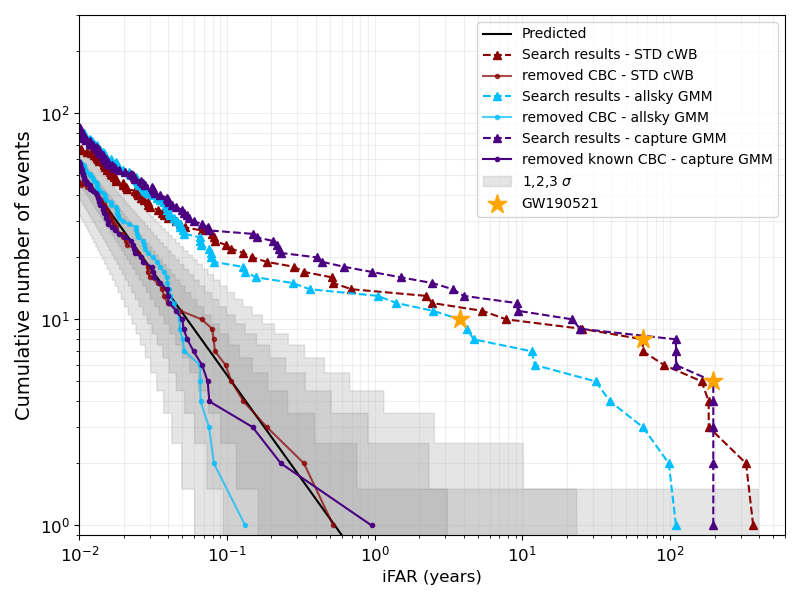}
    \caption{Comparison of search results on the combined O3 data with the capture-informed GMM analysis (purple), all-sky short GMM analysis (blue) and the standard cWB analysis (red). Full search results are represented by triangular dashed markers, and include known CBC events. Results with known CBC events removed are represented by solid circular markers, all consistent with the expected background. The GW190521 event is marked by a star within each of the searches.}
    \label{fig:openbox_all}
\end{figure}

Given the increased sensitivity to dynamical captures observed with the informed GMM approach, we analyze the O3 detector data with the same approach to search for significant GW events, presenting the search results in Figure \ref{fig:openbox_all}. O3 data has previously been analysed by the generic all-sky short GMM and standard cWB analyses in \cite{Smith:2024gmm} and \cite{allskyo3} respectively, thus their results will not be discussed in detail here, however they are included in Figure \ref{fig:openbox_all} for comparison. In the informed-GMM search, all detected GW transients detected with significance iFAR $\geq$ 1 year are established CBC candidates detailed in \cite{GWTC-2,GWTC-3}. Once known CBC events are removed, search results are consistent with the expected background, concluding with a null result on new detections.  

The loudest remaining trigger in the informed GMM search after known CBCs are removed is found at a UTC time of 2019-08-11 13:03:13 with a significance of iFAR = 0.95 years. The trigger was observed at a lower significance of iFAR = 0.008 years in both the all-sky short GMM search and standard cWB search. The event has a network signal-to-noise ratio (SNR) of 5.92 concentrated around 118 Hz. 
No apparent glitches or data quality issues are observed around the time of the event, thus it cannot be immediately discarded on the grounds of data quality. 
Due to the detected significance being slightly under the 1 year threshold and the trigger lying within the 1 sigma region of the expected background, we conclude that the probable origin of this event is a noise artifact.

\begin{table}
    \centering
    \setlength{\tabcolsep}{7pt}
    \begin{tabular}{c | c c c }
         \hline
         \hline

            \multirow{3}{*}{Event Name}  & Informed   & All-sky  &  Standard   \\
               & GMM  & GMM  & cWB  \\
              & iFAR [yr]  & iFAR [yr]  & iFAR [yr]   \\
          \hline

          GW190521 & > 196.19 & 3.77 &  65.38 \\
          GW190706\_222641 & > 196.19 & 4.67 & 65.38    \\
          GW190521\_074359 & > 196.19 & 98.09 & 326.88 \\
          GW190412 & > 196.19 & 65.40 & 15.10   \\
          GW190519\_153544  & > 196.19 & 39.24 & 7.78  \\
          GW200224\_222234 & 109.62 & 109.62 & 365.32   \\
          GW191109\_010717 & 109.62 & 12.18 & 182.66  \\
          GW200311\_115853 & 109.62 & 4.22 & 182.66  \\
          GW190828\_063405 & 24.52 & 11.54 & 163.44  \\
          GW190602\_175927 & 21.80 & 0.28 & 0.51 \\
          GW190915\_235702 & 9.34 & 1.05 & 5.36 \\
          GW191222\_033537 & 9.14 & 1.49 & 2.45  \\
          GW190421\_213856 & 4.00 & 0.08 & 0.05 \\
          GW190408\_181802 & 3.38 & 2.48 & 25.15  \\
          GW200225\_060421 & 2.46 & 1.06 & 2.45  \\
          GW190503\_185404 & 1.50 & 0.13 & 0.70 \\
          GW191204\_171526 & 0.06 & 31.32 & 91.33 \\

         \hline
         \hline
    \end{tabular}
    \caption{Table showing observed CBC events with iFAR $\geq$ 1 year over all 3 search methods. Events are listed in decreasing significance of the informed-GMM approach. Events which have a `>' symbol preceding their iFAR estimate are observed with the highest possible significance considering our available background.}
    \label{detection_table}
\end{table}

A total of 16 known CBC events are detected at iFAR $\geq$ 1 year with the capture-informed GMM, 5 of which are detected with the maximum significance for our available background. The number of detected events is more than seen by the all-sky-short GMM approach (14) \cite{Smith:2024gmm} and the standard cWB approach (14) \cite{allskyo3}. 3 CBC events observed by the informed GMM analysis were previously missed by the all-sky GMM and standard cWB analyses, namely; GW190602 detected with iFAR = 21.80 years and total mass 116.3 $M_{\odot}$, GW190421 detected with iFAR = 4.00 years and total mass 72.9 $M_{\odot}$, and GW190503 detected with iFAR = 1.50 years and total mass 71.7 $M_{\odot}$ \cite{GWTC-2}.  
GW191204 is missed by the informed GMM analysis however is detected by both other approaches, with a total mass of 20.2M$_{\odot}$.
Of the 13 common detections, the informed GMM analysis detects 13 (10) events with increased significance compared to the all-sky short GMM (standard cWB) analysis.
The increased performance of the capture-informed GMM on CBC events is not surprising, given that the model is trained on a sub-population of CBC systems. In particular, detection significance appears to increase for higher mass CBC systems, likely due to the similar morphology between dynamical capture and high mass quasi-circular binaries \cite{Guo:2022mimick}. 

One of the highest significance CBC detections from the informed GMM search is GW190521, labeled as a probable IMBH event in previous analysis \cite{GWTC-2,gw190521-discovery,gw190521-properties}. However work by Gamba et al. \cite{Gamba:2023capture} suggests that this event may arise from dynamical capture, making it of interest in our study. The informed GMM search detects GW190521 at an iFAR of 196.19 years, the highest available significance estimate from our considered background, compared to a detection significance of 3.77 years in the all-sky short GMM search and 65.38 years in the standard cWB search.

\subsection{Upper limits on expected rates}

We extend the result of our sensitivity study to estimate upper limits on the observational rate of dynamical black hole capture events. 
Given that the generic all-sky short GMM approach does not find significantly improved sensitivities compared to the previous work, we consider the rate calculation only for the informed GMM approach. We consider two hypotheses for the rate calculation; that GW190521 is not a dynamical capture event, thus making the number of capture events detected zero, and that GW190521 is in fact a dynamical capture event, as presented in \cite{Gamba:2023capture}. 

First we consider the hypothesis where no dynamical capture events are detected. Assuming the population of dynamical capture sources produces events within a Poisson distribution with rate $R$, we can compute the upper limit on event rate at a 90\% confidence interval given the non-detection of events over a given iFAR $x$ through \cite{Brady:200upp-rate}:

\begin{equation}
    R_{90} = \frac{2.303}{\langle VT \rangle _{i}^{\text{iFAR}>x}}
\end{equation}
Here we consider an iFAR threshold of $x=100$ years for detection. 

\begin{figure}
    \centering
    \includegraphics[width=\linewidth]{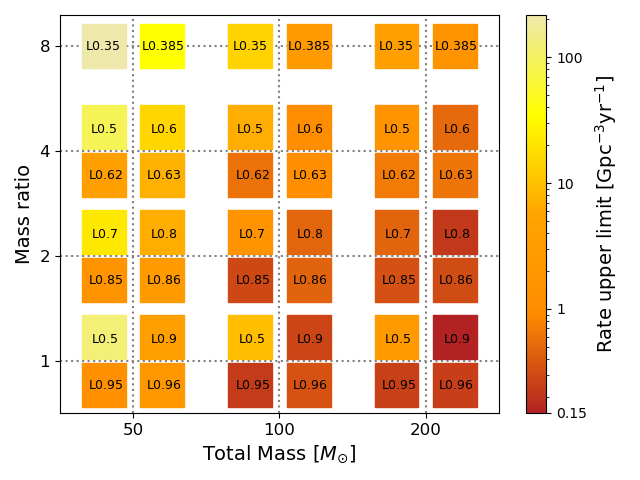}
    \caption{ Estimates of rate upper limit for each considered waveform based on the null detection of dynamical capture events with the informed GMM model. The initial angular momentum of each waveform is displayed on the square markers, while the rate estimate is represented by the color bar. More stringent rate estimates are in red. }
    \label{fig:rates}
\end{figure}

The rate upper limits achieved with the capture-informed GMM search are presented in Fig. \ref{fig:rates}. Results for 20 $M_{\odot}$ total mass waveforms are omitted from the figure due to lower sensitivity compared to other waveforms, however can be found in Table \ref{tab:comparison}, alongside the results reported in our previous work.
The most stringent rate upper limit at 90\% confidence is placed on the $q=1, L=0.9 $, $M_{tot}$=200 $M_{\odot}$ waveform at 0.15 Gpc$^{-3}$yr$^{-1}$, a 34\% improvement on the previously quoted upper limit observational rate under a search with minimal assumptions \cite{Ebersold:2022capture}. 
From literature, it is expected for single-single captures following a similar astrophysical model as considered here to occur at 0.002-0.04 Gpc$^{-3}$yr$^{-1}$ \cite{Rasskazov:2019s-s}. As can be seen, our rate upper limit estimate is at least an order of magnitude too large to be competitive with quoted rates, however it is probable that such rates may be achieved in the next LVK observing runs. 

\begin{table*}
    \centering
    \begin{tabular}{|c|c|c|c|c c c|c c|  |c|c|c|c|c c c|c c|}
         \hline
         \hline

          $M_{tot}$ & $q$ & $L$ & $z_{max}$ & \multicolumn{3}{c|}{$D_{\langle VT \rangle}$} & \multicolumn{2}{c||}{$R_{90}$} & $M_{tot}$ & q & L & $z_{max}$ & \multicolumn{3}{c|}{$D_{\langle VT \rangle}$} & \multicolumn{2}{c|}{$R_{90}$}  \\
          $[M_{\odot}]$ &  &  &  & \multicolumn{3}{c|}{[Gpc]} & \multicolumn{2}{c||}{[Gpc$^{-3}$yr$^{-1}$]}  & $[M_{\odot}]$ &  &  &  & \multicolumn{3}{c|}{[Gpc]} & \multicolumn{2}{c|}{[Gpc$^{-3}$yr$^{-1}$]}  \\
          \hline
            &  & & & Capture & all-sky & STD & Capture & STD & & & & & Capture & all-sky & STD  & Capture & STD  \\
           &  &  &  &  GMM &  GMM &  cWB & GMM & cWB & & & & &  GMM &  GMM &  cWB & GMM & cWB \\
          \hline\hline
          \multirow{14}{*}{20} & \multirow{4}{*}{1} & 0.5 & 0.03  & 0.048 & 0.036 & 0.030 & 8930 & 37641 & \multirow{14}{*}{50} & \multirow{4}{*}{1} & 0.5 &  0.15 & 0.20 & 0.16 & 0.12  & 126 & 583 \\
          %\cline{3-7}\cline{10-14}
            &  &  0.9 & 0.08 & 0.12 & 0.089 & 0.10 & 657 & 1002 &  &  &  0.9 & 0.45  & 0.64 & 0.49 & 0.58  & 3.86 & 5.22 \\
          %\cline{3-7}\cline{10-14}
            & &  0.95 & 0.15  & 0.19 & 0.15 & 0.17  & 140.5 & 190.4 &  & &  0.95 &  0.7 & 0.93 & 0.71 & 0.82  & 1.25 & 1.82 \\
          %\cline{3-7}\cline{10-14}
            & & 0.96 & 0.15 & 0.20 & 0.16 & 0.19  & 124.4 & 145.2 &   & & 0.96 & 0.65 & 0.80 & 0.65 & 0.78  & 2.01 & 2.15 \\
          \cline{2-9}\cline{11-18}
          %%%%%%
           & \multirow{4}{*}{2} & 0.7 & 0.04  & 0.071 & 0.055 & 0.050  & 2849 & 8074 & & \multirow{4}{*}{2} & 0.7 &  0.3 & 0.36 & 0.28 & 0.26  & 22.0 & 55.3 \\
          %\cline{3-7}\cline{10-14}
           & &  0.8 & 0.06 & 0.10 & 0.077 & 0.085  & 899.4 & 1625 &  & &  0.8 & 0.4 & 0.53 & 0.41 & 0.47  & 6.80 & 9.99 \\
          %\cline{3-7}\cline{10-14}
           &  & 0.85 & 0.12  & 0.18 & 0.14 & 0.16  & 162 & 248.6 &  &  & 0.85 & 0.56 & 0.84 & 0.64 & 0.74  & 1.68 & 2.47 \\
          %\cline{3-7}\cline{10-14}
            & & 0.86 & 0.12 & 0.17 & 0.14 & 0.16  & 193.1 & 241.3 &   & & 0.86 & 0.56  & 0.72 & 0.57 & 0.69  & 2.74 & 3.05 \\
          \cline{2-9}\cline{11-18}
          %%%%%
           & \multirow{4}{*}{4} & 0.5 & 0.03  & 0.049 & 0.038 & 0.032  & 8371 & 28715 & & \multirow{4}{*}{4} & 0.5 & 0.15 & 0.22 & 0.18 & 0.16  & 89.36 & 268 \\
          %\cline{3-7}\cline{10-14}
           & & 0.6  & 0.05 & 0.082 & 0.062 & 0.071  & 1832 & 2779 & & & 0.6  & 0.25 & 0.41 & 0.32 & 0.36  & 15.08 & 21.6 \\
          %\cline{3-7}\cline{10-14}
            & &  0.62 & 0.09  & 0.13 & 0.11 & 0.12  & 450.7 & 581 &  & &  0.62 & 0.4 & 0.64 & 0.48 & 0.55  & 3.94 & 6.08 \\
          %\cline{3-7}\cline{10-14}
            & & 0.63  & 0.08 & 0.12 & 0.097 & 0.11  & 662.8 & 690.6 &  & & 0.63 & 0.4 & 0.52 & 0.42 & 0.52  & 7.15 & 7.41 \\
          \cline{2-9}\cline{11-18}
          %%%%%
           & \multirow{2}{*}{8} & 0.35 & 0.025 & 0.039 & 0.030 & 0.028  & 17228 & 48546 & & \multirow{2}{*}{8} & 0.35 & 0.12 & 0.17 & 0.13 & 0.13  & 216.39 & 500 \\
          %\cline{3-7}\cline{10-14}
           & &  0.385 & 0.04 & 0.062 & 0.049 & 0.059  & 4169 & 4811 &  & &  0.385 & 0.2 & 0.31 & 0.24 & 0.28  & 34.73 & 45.9 \\
          \hline\hline
          \multirow{14}{*}{100} & \multirow{4}{*}{1} & 0.5 & 0.32 & 0.47 & 0.38 & 0.35  & 9.44 & 24.5 & \multirow{14}{*}{200} & \multirow{4}{*}{1} & 0.5 & 0.42 & 0.76 & 0.61 & 0.53  & 2.31 & 6.79 \\
          %\cline{3-7}\cline{10-14}
           & &  0.9 & 1.02 & 1.54 & 1.18 & 1.31  & 0.28 & 0.45 &  & &  0.9 & 1.2 & 1.90 & 1.55 & 1.64  & 0.15 & 0.23 \\
          %\cline{3-7}\cline{10-14}
           &  & 0.95 & 1.12  & 1.63 & 1.35 & 1.50  & 0.23 & 0.30 &  &  & 0.95 & 1.16 & 1.60 & 1.35 & 1.46  & 0.25 & 0.32 \\
          %\cline{3-7}\cline{10-14}
           &  & 0.96 & 1.07 & 1.44 & 1.16 & 1.35  & 0.34 & 0.41 & &  & 0.96 & 1.16 & 1.61 & 1.31 & 1.38  & 0.24 & 0.38 \\
          \cline{2-9}\cline{11-18}
          %%%%%
           & \multirow{4}{*}{2} & 0.7 & 0.8  & 0.87 & 0.66 & 0.67  & 1.52 & 3.43 & & \multirow{4}{*}{2} & 0.7 & 0.9 & 1.29 & 1.01 & 0.98  & 0.48 & 1.08 \\
          %\cline{3-7}\cline{10-14}
           &  &  0.8 & 0.93 & 1.28 & 0.99 & 1.05  & 0.48 & 0.87 &  &  &  0.8 & 1.0 & 1.66 & 1.33 & 1.39  & 0.22 & 0.37 \\
          %\cline{3-7}\cline{10-14}
            & & 0.85 & 1.03  & 1.52 & 1.24 & 1.37  & 0.29 & 0.39 &  & & 0.85 & 0.92 & 1.44 & 1.22 & 1.33  & 0.34 & 0.43 \\
          %\cline{3-7}\cline{10-14}
            & & 0.86 & 0.93 & 1.31 & 1.06 & 1.23  & 0.45 & 0.54 & & & 0.86 &  0.92 & 1.49 & 1.19 & 1.26  & 0.30 & 0.51 \\
          \cline{2-9}\cline{11-18}
          %%%%
           & \multirow{4}{*}{4} & 0.5 &  0.32 & 0.54 & 0.43 & 0.41  & 6.57 & 15.1 & & \multirow{4}{*}{4} & 0.5 & 0.41 & 0.86 & 0.67 & 0.60  & 1.60 & 4.60 \\
          %\cline{3-7}\cline{10-14}
            & &  0.6 & 0.62 & 1.00 & 0.75 & 0.79  & 1.00 & 2.02 &  & &  0.6 & 0.79 & 1.26 & 0.99 & 1.02  & 0.51 & 0.95 \\
          %\cline{3-7}\cline{10-14}
            & & 0.62 & 0.77  & 1.20 & 0.98 & 1.07  & 0.59 & 0.83 &  & & 0.62 & 1.07 & 1.14 & 0.96 & 1.03  & 0.68 & 0.92 \\
          %\cline{3-7}\cline{10-14}
            & & 0.63 & 0.66 & 0.99 & 0.80 & 0.90  & 1.03 & 1.40 &   & & 0.63 & 0.9 & 1.18 & 0.94 & 0.98  & 0.62 & 1.09 \\
          \cline{2-9}\cline{11-18}
          %%%%%
           & \multirow{2}{*}{8} & 0.35 &  0.28 & 0.41 & 0.33 & 0.31  & 14.48 & 33.30 & & \multirow{2}{*}{8} & 0.35 & 0.31 & 0.63 & 0.49 & 0.44  & 4.11 & 11.6 \\
          %\cline{3-7}\cline{10-14}
            & & 0.385  & 0.42 & 0.74 & 0.56 & 0.63  & 2.48 & 4.03 &  & & 0.385  & 0.63 & 0.84 & 0.66 & 0.72  & 1.70 & 2.70 \\
          \hline\hline

    \end{tabular} 
    \caption{Table displaying values for sensitive distance and rate upper limit at iFAR $\geq$ 100 years for the considered GMM approaches, with comparison to the previous results from \cite{Ebersold:2022capture}. Rate upper limit is not considered for the all-sky short GMM since sensitivity estimates were not competitive with the other methods.  }
    \label{tab:comparison}
\end{table*}

While the above rate upper limit calculations assume a null detection on dynamical capture events, work by Gamba et al. \cite{Gamba:2023capture} suggests that the detected event GW190521 may be a dynamical capture event with total mass 130$^{+75}_{-43}$ $M_{\odot}$ and mass ratio $q\sim 1$ or $q\sim 2$. Due to the ambiguity on the origin of this event, we perform the sensitive distance and rate upper limit calculations assuming it is indeed a dynamical capture event, detected with the informed GMM analysis at a significance of iFAR $\geq$ 196 years. We consider injections of $M_{tot}$=[100,200] $M_{\odot}$ and $q=[1,2]$. 
The rate upper limit considering one detection at iFAR of 196 years is calculated as \cite{Brady:200upp-rate}:

\begin{equation}
    R_{90} = \frac{3.9}{\langle VT \rangle _{i}^{\text{iFAR}>196yr}}
\end{equation}

The sensitive distance and rate upper limit results at iFAR $\geq$ 196 years are shown in Table \ref{tab:rates_gw190521}. 
In the scenario where GW190521 is a dynamical capture event, the most stringent upper limit on the expected rate at 90\% confidence is 1.73 Gpc$^{-3}$yr$^{-1}$ for 100 $M_{\odot}$ and 0.94 Gpc$^{-3}$yr$^{-1}$ for 200 $M_{\odot}$. The estimated upper limits on rate are higher than quoted in the null detection hypothesis due to lower sensitive volume at the detection significance of GW190521.

\begin{table}
    \centering
    \setlength{\tabcolsep}{8pt}
    \begin{tabular}{|c|c|c|c|c|c|}
        \hline
        \hline
        $M_{tot}$ & $q$ & $L$ & $z_{max}$ & $D_{\langle VT \rangle}$ & $R_{90}$ \\
        $[M_{\odot}]$ &   &   &   & [Gpc] & [Gpc$^{-3}$yr$^{-1}$] \\
        \hline
        \multirow{8}{*}{100} & \multirow{4}{*}{1}  & 0.5  & 0.32  & 0.27 & 86.4 \\
          &   & 0.9  & 1.02 & 0.86 & 2.69 \\
          &   & 0.95 & 1.12  & 1.00 & 1.73 \\
          &   & 0.96 & 1.07 & 0.87 & 2.61 \\
        \cline{2-6}
          & \multirow{4}{*}{2}  &  0.7 &  0.8 & 0.46 & 17.2 \\
          &   & 0.8 & 0.93 & 0.67 & 5.67\\
          &   & 0.85 & 1.03 & 0.90 & 2.33 \\
          &   & 0.86 & 0.93 & 0.80 & 3.38 \\
          \hline
          \multirow{8}{*}{200} & \multirow{4}{*}{1}  & 0.5  & 0.32  & 0.45 & 19.0 \\
          &   & 0.9  & 1.02 & 1.22 & 0.94 \\
          &   & 0.95 & 1.12  & 1.03 & 1.58 \\
          &   & 0.96 & 1.07 & 1.02 & 1.62 \\
        \cline{2-6}
          & \multirow{4}{*}{2}  &  0.7 &  0.8 & 0.78 & 3.56 \\
          &   & 0.8 & 0.93 & 1.04 & 1.52 \\
          &   & 0.85 & 1.03 & 0.92 & 2.19 \\
          &   & 0.86 & 0.93 & 0.94 & 2.05 \\
          \hline
          \hline
    \end{tabular}
    \caption{Sensitive distance and rate upper limit estimates for injections with mass and mass ratio close to the estimated properties of GW190521 as a dynamical capture, at an iFAR threshold of 196 years. The values are shown only for the informed GMM analysis.}
    \label{tab:rates_gw190521}
\end{table}

\section{Conclusions}
\label{Sec:conclusions}

Dynamical captures of black holes are expected to emit GWs within the sensitive frequency-band of current ground-based detectors. In this work, we conducted a search for such events using O3 LVK data, building upon our previous efforts by applying two configurations of the enhanced cWB+GMM algorithm, including the first application of an informed-GMM analysis to target a specific astrophysical population. 

While no new significant event was identified in the search, the informed-GMM analysis led to an increased detection significance for known CBC signals compared to previous cWB results. In particular, GW190521 was observed at an increased significance of iFAR $\geq$ 196.19 years. The ambiguous nature of this event and its potential origin from dynamical capture highlights the advantage an informed approach brings to the detection of such sources.  

The benefit of an informed GMM post-processing was further demonstrated through the sensitivity study performed on the numerical relativity capture injections. The observable sensitive distance of GW dynamical capture events has been increased to 1.9 Gpc, enabling a 34\% improvement on the estimated rate upper limits assuming null detection. Rate upper limit estimates were also calculated assuming GW190521 was a dynamical capture event, at $R_{90}\sim 0.94 - 1.73$ Gpc$^{-3}$yr$^{-1}$.

Despite the slight loss in sensitivity observed for some waveforms with the cWB-GMM weakly-modeled all-sky short approach compared to previous efforts, sensitivity to low angular momentum waveforms is significantly increased due to the mitigation of blip glitches. Additionally, it remains possible to detect GWs from the dynamical capture of black holes up to 1 Gpc with such a search.

Looking ahead to future observing runs, the development of search techniques for dynamical capture events becomes increasingly important due to increased detector sensitivity, particularly in the low frequency region where such events are expected to occur. Future work will include a reinforced training set to expand the coverage of the dynamical capture parameter space, as well as extending the search to other waveform variations, such as spinning black hole capture and repeated interactions before merger. Moreover, the informed GMM analysis developed here holds promise for applications to other source populations, potentially aiding in the detection of new GW events.

\begin{acknowledgments}
The authors would like to thank Gayathri V and Giovanni Prodi for comments. 
This material is based upon work supported by NSF's LIGO Laboratory which is a major facility fully funded by the National Science Foundation. The authors acknowledge the computational resources which aided the completion of this project, provided by LIGO-Laboratory and supported by the National Science Foundation (NSF) Grants No..PHY-0757058 and No.PHY-0823459.
L.S acknowledges support from the European Union - Next Generation EU Mission 4 Component 1 CUP J53D23001550006 with the PRIN Project No. 202275HT58, and by ICSC – Centro Nazionale di Ricerca in High Performance Computing, Big Data and Quantum Computing, funded by European Union – NextGenerationEU. ST is supported by the Swiss National Science Foundation Ambizione Grant Number: PZ00P2-202204. Y.-B.B. was supported by the National Research Foundation of Korea (NRF) grant funded by the Korea government(MSIT) RS-2025-00556091. ISH was supported by the Science and Technology Facilities Council (STFC) grants ST/V001736/1 and ST/V005634/1.

\end{acknowledgments}

\appendix

\section{Selection of cWB attributes}\label{app:attributes}

The GMM analysis requires the selection of cWB attributes which represent the distributions of noise and signal well in the multi-dimensional parameter space. As detailed in \cite{Smith:2024gmm}, the cWB-GMM analysis to the all-sky short search considers 11 cWB attributes: 
effective network coherent SNR ($\eta_{c}$), network correlation coefficients ($c_{c0}$, $c_{c2}$), the network coherent energy ($E_{c}$), the network energy dis-balance ($N_{ED}$), the ratio between the reconstructed energy and the total energy ($N_{\mathrm{norm}}$), a penalty factor on the energy content of the reconstructed trigger (penalty), and attributes measuring likeness to known glitches ($Q_{veto0,1}$, $L_{veto0,1}$). 

These attributes are re-parameterized in order to provide better Gaussian behavior. A list of the cWB attributes and their corresponding re-parameterisation is seen in Table \ref{tab:attributes}.

\begin{table}
    \centering
    \setlength{\tabcolsep}{10pt}
    \begin{tabular}{c | c}
         \hline
         \hline

          Original attribute & Re-parameterized attribute  \\ 
             &  LH  \\[1ex] 
          \hline
 
          $E_{c}$ &  $\log_{10}(E_{c})$ \\
          $\eta_{c}$ &  $\log_{10}(\eta_{c})$ \\ 
          $c_{c0}$ & $\mathrm{logit}(c_{c0})$ \\
          $c_{c2}$ & $\mathrm{logit}(c_{c2})$ \\
          $N_{ED}$ & $\log_{10}(N_{ED}+1000)$\\
          $N_{\mathrm{norm}}$ & $N_{\mathrm{norm}}$ \\
          $\mathrm{penalty}$ & $\log_{10}(\mathrm{penalty})$ \\ 
          $Q_{\mathrm{veto0}}$ & $\log_{10}(Q_{\mathrm{veto0}}+1)$ \\
          $Q_{\mathrm{veto1}}$ & $\log_{10}(Q_{\mathrm{veto1}})$ \\
          $L_{\mathrm{veto0}}$ & \multirow{2}{*}{$\mathrm{logit}(L_{\mathrm{ratio}}) = \mathrm{logit}(\frac{L_{\mathrm{veto1}}}{L_{\mathrm{veto0}}})$} \\
          $L_{\mathrm{veto1}}$ & \\[1ex] 
    \hline
    \hline
         
    \end{tabular}
    \caption{Table of cWB attributes selected for GMM analysis and their re-parameterisation. }
    \label{tab:attributes}
\end{table}

\bibliographystyle{apsrev4-1}
\bibliography{references}

\end{document}